\documentstyle[sprocl]{article}
\input{psfig}
\def\be{\begin{equation}}
\def\ee{\end{equation}}
\def\bea{\begin{eqnarray}}
\def\eea{\end{eqnarray}}

\def\VEV#1{\left\langle #1\right\rangle}
\def\fun#1#2{\lower3.6pt\vbox{\baselineskip0pt\lineskip.9pt
  \ialign{$\mathsurround=0pt#1\hfil##\hfil$\crcr#2\crcr\sim\crcr}}}
\def\la{\mathrel{\mathpalette\fun <}}
\def\ga{\mathrel{\mathpalette\fun >}}
\def\gtrsim{\ga}
\def\lesssim{\la}
\def\slashchar#1{\setbox0=\hbox{$#1$}           
   \dimen0=\wd0                                 
   \setbox1=\hbox{/} \dimen1=\wd1               
   \ifdim\dimen0>\dimen1                        
      \rlap{\hbox to \dimen0{\hfil/\hfil}}      
      #1                                        
   \else					
      \rlap{\hbox to \dimen1{\hfil$#1$\hfil}}   
      /                                         
   \fi}                                         %

\begin{document}

\rightline{CU-TP-866, CAL-648, hep-ph/9710467}
\vskip 0.5in
\title{WIMP AND AXION DARK MATTER\footnote{To appear in the
proceedings of the 1997 ICTP Summer School on High Energy
Physics and Cosmology, Trieste, Italy, June 2--July 4, 1997.}}
\author{M. KAMIONKOWSKI}
\address{Department of Physics, Columbia University, 538 West
120th Street, New York, NY 10027 USA}
%
%

\maketitle
\abstracts{
There is almost universal agreement among cosmologists that most
of the matter in the Universe is dark, and there are very good
reasons to believe that most of this dark matter must be nonbaryonic.
The two leading candidates for this dark matter are axions and 
weakly-interacting massive particles (WIMPs), such as the
neutralino in supersymmetric extensions of the standard model.
I discuss the arguments for these two dark-matter candidates and
review techniques for discovery of these dark-matter particles.}

\section{Introduction}
Almost all astronomers will agree that most of the mass in the
Universe is nonluminous.  The nature of this dark matter remains
one of the great mysteries of science today.  
Dynamics of cluster of galaxies suggest
a universal nonrelativistic-matter density of
$\Omega_0\simeq0.1-0.3$.  If the luminous matter were all
there was, the duration of the epoch
of structure formation would be very short, thereby requiring
(in almost all theories of structure formation) fluctuations in
the microwave background which would be larger than those observed.  
These considerations imply $\Omega_0\gtrsim0.3$~\cite{marccmb}.
Second, if the current value of $\Omega_0$ is of order unity today,
then at the Planck time it must have
been $1\pm10^{-60}$ leading us to believe that $\Omega_0$ is
precisely unity
for aesthetic reasons.  A related argument comes from
inflationary cosmology, which provides the most satisfying
explanation for the smoothness of the microwave
background~\cite{inflation}.  To account for this isotropy,
inflation must set $\Omega$ (the {\it total} density, including
a cosmological constant) to unity.

\begin{figure}[htbp]
\centerline{\psfig{file=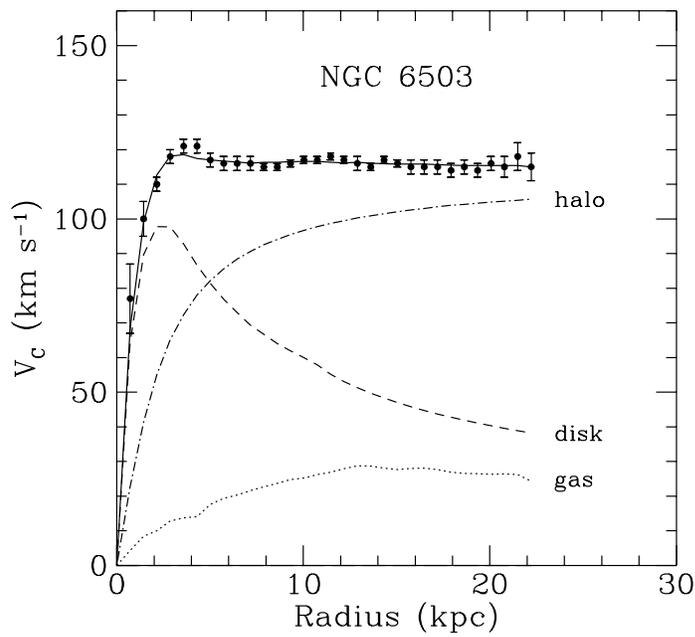,width=3.7in}}
\bigskip
\caption{\baselineskip=0.4cm
         Rotation curve for the spiral galaxy NGC6503.  The points
	 are the measured circular rotation velocities as a
	 function of distance from the center of the galaxy.
	 The dashed and dotted curves are the contribution to
	 the rotational velocity due to the observed disk and
	 gas, respectively, and the dot-dash curve is the
	 contribution from the dark halo.}
\label{rotationfigure}
\end{figure}

However, the most robust observational evidence for the
existence of dark matter involves galactic dynamics.  There is
simply not enough luminous matter ($\Omega_{\rm lum}\la0.01$)
observed in spiral galaxies to account for their observed
rotation curves (for example, that for NGC6503 shown in
Fig.~\ref{rotationfigure}~\cite{broeils}).  Newton's
laws imply galactic dark halos with masses that contribute
$\Omega_{\rm halo} \ga 0.1$.

On the other hand, big-bang nucleosynthesis
suggests that the baryon density is $\Omega_b\la0.1$~\cite{bbn}, too
small to account for the dark matter in the Universe.
Although a neutrino species of mass ${\cal O}(30\, {\rm eV})$ could
provide the right dark-matter density, N-body simulations of
structure formation in a neutrino-dominated Universe do a poor
job of reproducing the observed structure~\cite{Nbody}.
Furthermore, it is difficult to see (essentially from the Pauli
principle) how such a neutrino could make up the dark matter in
the halos of galaxies~\cite{gunn}. It appears likely then, that some
nonbaryonic, nonrelativistic matter is required.

The two leading candidates from particle theory are the
axion~\cite{axion}, which arises in the Peccei-Quinn solution to
the strong-$CP$ problem, and a weakly-interacting massive particle
(WIMP), which may arise in supersymmetric (or other) extensions
of the standard model~\cite{jkg}.  As discussed below, there are
good reasons to believe that if the Peccei-Quinn mechanism is
responsible for preserving $CP$ in the strong
interactions, then the axion is the dark matter.  Similarly,
there are also excellent reasons to expect that if low-energy
supersymmetry exists in Nature, then the dark matter should be
composed of the lightest superpartner.  The study of these ideas
are no longer exclusively the domain of theorists: there are now
a number of experiments aimed at discovery of axions and WIMPs.
If axions populate the Galactic halo, they can be converted to
photons in resonant cavities immersed in strong magnetic
fields.  An experiment to search for axions in this fashion is
currently being carried out.  If WIMPs populate the halo, they
can be detected either directly in low-background laboratory
detectors or indirectly via observation of energetic neutrinos
from WIMPs which have accumulated and then annihilated in the
Sun and/or Earth.

Here, I first show how the observed dynamics of the Milky Way
indicate a local dark-matter density of $\rho_0\simeq0.4$
GeV~cm$^{-3}$.  I then review the arguments for axion and WIMP dark matter and
the methods of detection.  However, there is no way I can do
this very active field of research justice in such short space.
For background on early-Universe cosmology, I recommend the book
by Kolb and Turner~\cite{kolbturnerbook}.  Readers with further
interest in WIMPs should see the recent review article by
Jungman, Griest and me~\cite{jkg}.  The first four sections of
that article are meant to provide a general review of dark and
supersymmetry, and  the idea of WIMP dark matter.  The remainder
of that article provides technical
details required by those interested in actively pursuing
research on the topic.  There are also several excellent axion
reviews~\cite{axion} and the recent book by Raffelt~\cite{raffelt}.

\section{The Local Dark-Matter Density}

The extent of the luminous disk of our Galaxy, the Milky Way, is
roughly 10 kpc, and we live about 8.5 kpc from the center.
Due to our location {\it in} it, the rotation curve of the Milky
Way cannot be determined with the same precision as that of an
external spiral galaxy, such as that shown in
Fig. \ref{rotationfigure}.  However, it is qualitatively the
same.  The circular speed rises linearly from zero at the center
and asymptotes to roughly 220 km~sec$^{-1}$ somewhere near our
own Galactocentric radius and remains roughly flat all the way
out to $\sim25$~kpc.  Although direct measurements of the
rotation curve are increasingly difficult at larger radii, the
orbital motions of satellites of the Milky Way suggest that the
rotation curve remains constant all the way out to radii of 50
kpc and perhaps even farther.  According to Newton's laws, the
rotation speed should fall as $v_c \propto r^{-1/2}$ at radii
greater than the extent of the luminous disk.  However, it is
observed to remain flat to much larger distances.  It therefore
follows that the luminous disk and bulge must be immersed
in an extended dark halo (or that Newton's laws are violated).

Our knowledge of the halo comes almost solely from this rotation
curve.  Therefore, we do not know empirically if the halo is
round, elliptical, or perhaps flattened like the disk.  However,
there are good reasons to believe that the halo should be much
more diffuse than the disk.  The disk is believed to be flat
because luminous matter can radiate photons and therefore
gravitationally collapse to a pancake-like structure.  On the
other hand, dark matter (by definition) cannot radiate photons.
There are also now empirical arguments which involve, e.g., the
shape of the distribution of gas in the Milky Way, which suggest
that the dark halo should be much more diffuse than the disk \cite{rob}.

Given that the halo is therefore nearly round, it must have a
density distribution like
\begin{equation}
     \rho(r) = \rho_0 {r_0^2 + a^2 \over r^2 + a^2},
\label{eq:rho}
\end{equation}
where $r$ is the radius, $r_0\simeq8.5$~kpc is our distance from
the center, $a$ is the core radius of the halo, and $\rho_0$ is
the local halo density.  Such a halo would give rise to a
rotation curve,
\begin{equation}
     v_h^2(r) = 4\pi G \rho_0 (r_0^2 +a^2)
     [1-(a/r)\tan^{-1}(r/a)],
\end{equation}
where $G$ is Newton's constant.  If we know the rotation speed
contributed by the halo at two points, we can determine $\rho_0$
and $a$.  At large radii, the rotation curve of the Milky Way is
supported entirely by this dark halo, so $v_h(r\gg10\,{\rm kpc})
\simeq 220$~km~sec$^{-1}$.  However, the rotation curve locally
is due in part to the disk, $v_c^2(r_0) =
v_d^2(r_0)+v_h^2(r_0)$.  The disk contribution to the local
rotation speed is somewhat uncertain but probably falls in the
range $v_d(r_0) \simeq 118-155$~km~sec$^{-1}$, which means that the
halo contribution to the local rotation speed is $v_h(r_0)
\simeq 150-185$~km~sec$^{-1}$.  Given the local and asymptotic
rotation speeds, we infer that the local halo density is $\rho_0
\simeq 0.3-0.5$~GeV~cm$^{-3}$.  The particles which make up the
dark halo move locally in the same gravitational potential well
as the Sun.  Therefore, the virial theorem tells us that they
must move with velocities $v\sim v_c \sim220$~km~sec$^{-1}$.
Additional theoretical arguments suggest that the velocity
distribution of these particles is locally nearly isotropic and
nearly a Maxwell-Boltzmann distribution.  To sum, application of
Newton's laws to our Galaxy tells us that the luminous disk and
bulge must be immersed in a dark halo with a local density
$\rho_0 \simeq 0.4$~GeV~cm$^{-3}$ and that dark-matter particles
(whatever they are) move with velocities comparable to the local
circular speed.  More careful investigations along these lines
show that similar conclusions are reached even if we allow for
the possibility of a slightly flattened halo or a radial
distribution which differs from that in Eq. (\ref{eq:rho})~\cite{ali}.

\section{Axions}

Although supersymmetric particles seem to get more attention in
the literature lately, we should not forget that the axion also
provides a well-motivated and promising alternative dark-matter
candidate~\cite{axion}.  The QCD Lagrangian may be written
\begin{equation}
     {\cal L}_{QCD} = {\cal L}_{\rm pert} + \theta {g^2 \over 32
     \pi^2} G \widetilde{G},
\end{equation}
where the first term is the perturbative Lagrangian responsible
for the numerous phenomenological successes of QCD.  However,
the second term (where $G$ is the gluon field-strength tensor
and $\widetilde{G}$ is its dual), which is a consequence of
nonperturbative effects, violates $CP$.  However, we know
experimentally that $CP$ is not violated in the strong
interactions, or if it is, the level of strong-$CP$ violation is
tiny.  From constraints to the neutron electric-dipole moment,
$d_n \lesssim 10^{-25}$ e~cm, it can be inferred that $\theta
\lesssim 10^{-10}$.  But why is $\theta$ so small?  This is the
strong-$CP$ problem.

The axion arises in the Peccei-Quinn solution to the strong-$CP$
problem~\cite{PQ}, which close to twenty years after it was proposed still
seems to be the most promising solution.  The idea is to
introduce a global $U(1)_{PQ}$ symmetry broken at a scale
$f_{PQ}$, and $\theta$ becomes a dynamical field which is the
Nambu-Goldstone mode of this symmetry.  
At temperatures below the QCD phase transition,
nonperturbative quantum effects break explicitly the symmetry
and drive $\theta\rightarrow 0$.  The axion is the
pseudo-Nambu-Goldstone boson of this near-global symmetry.  Its
mass is $m_a \simeq\, {\rm eV}\,(10^7\, {\rm GeV}/ f_a)$, and its
coupling to ordinary matter is $\propto f_a^{-1}$.

{\it A priori}, the Peccei-Quinn solution works equally well for
any value of $f_a$ (although one would generically expect it to
be less than or of order the Planck scale).  However, a variety
of astrophysical observations and a few laboratory experiments
constrain the axion mass to be $m_a\sim10^{-4}$ eV, to within a
few orders of magnitude.  Smaller masses would lead to an
unacceptably large cosmological abundance.  Larger masses
are ruled out by a combination of constraints from supernova
1987A, globular clusters, laboratory experiments, and a search
for two-photon decays of relic axions~\cite{ted}.

One conceivable theoretical difficulty with this axion mass
comes from generic quantum-gravity arguments~\cite{gravity}.  For
$m_a\sim10^{-4}$ eV, the magnitude of the explicit symmetry
breaking is incredibly tiny compared with the PQ scale, so the
global symmetry, although broken, must be very close to exact.
There are physical arguments involving, for example, the
nonconservation of global charge in evaporation of a black hole
produced by collapse of an initial state with nonzero global
charge, which suggest that  global symmetries should be violated
to some extent in quantum gravity.  When one writes down a
reasonable {\it ansatz} for a term in a low-energy effective
Lagrangian which might arise from global-symmetry violation at
the Planck scale, the coupling of such a term is found to be
extraordinarily small (e.g., $\lesssim 10^{-55}$).  Of course,
we have at this point no predictive theory of quantum gravity,
and several mechanisms for forbidding these global-symmetry
violating terms have been proposed~\cite{solutions}.  Therefore,
these arguments by no means ``rule out'' the axion solution.
In fact, discovery of an axion would provide much needed clues
to the nature of Planck-scale physics.

Curiously enough, if the axion mass is in the relatively small viable
range, the relic density is $\Omega_a\sim1$ and may therefore
account for the halo dark matter.  Such axions would be produced
with zero momentum by a misalignment mechanism in the early
Universe and therefore act as cold dark matter.  During the process of
galaxy formation, these axions would fall into the Galactic
potential well and would therefore be present in our halo with a
velocity dispersion near 270 km~sec$^{-1}$.

Although the interaction of axions with ordinary matter is
extraordinarily weak, Sikivie proposed a very clever method of
detection of Galactic axions~\cite{sikivie}.  Just as the axion couples to
gluons through the anomaly (i.e., the $G\widetilde{G}$ term),
there is a very weak coupling of an axion to photons through the
anomaly.  The axion can therefore decay to two
photons, but the lifetime is $\tau_{a\rightarrow \gamma\gamma}
\sim 10^{50}\, {\rm s}\, (m_a / 10^{-5}\, {\rm eV})^{-5}$ which
is huge compared to the lifetime of the Universe and therefore
unobservable.  However, the $a\gamma\gamma$ term in the
Lagrangian is ${\cal L}_{a\gamma\gamma} \propto a {\vec E} \cdot
{\vec B}$ where ${\vec E}$ and ${\vec B}$ are the electric and
magnetic field strengths.  Therefore, if one immerses a resonant
cavity in a strong magnetic field, Galactic axions which pass
through the detector may be converted to fundamental excitations
of the cavity, and these may be observable~\cite{sikivie}.  Such
an experiment is currently underway~\cite{axionexperiment}.
They have already begun to probe part of the cosmologically
interesting parameter space (no, they haven't found anything
yet) and expect to cover most of the interesting region
parameter space in the next three years.  A related experiment,
which looks for 
excitations of Rydberg atoms, may also find dark-matter
axions~\cite{rydberg}. 
Although the sensitivity of this technique should be
excellent, it can only cover a limited axion-mass range.

It should be kept in mind that there are no accelerator tests
for axions in the acceptable mass range.  Therefore, these
dark-matter axion experiment are actually our {\it only}
way to test the Peccei-Quinn solution.

\section{Weakly-Interacting Massive Particles}

Suppose that in addition to the known particles of the
standard model, there exists a new, yet undiscovered, stable (or
long-lived) weakly-interacting massive
particle (WIMP), $\chi$.  At temperatures
greater than the mass of the particle, $T\gg m_\chi$, the
equilibrium number density of such particles is $n_\chi \propto
T^3$, but for lower temperatures, $T\ll m_\chi$, the equilibrium
abundance is exponentially suppressed, $n_\chi \propto
e^{-m_\chi/T}$.  If the expansion of the Universe were so slow
that  thermal equilibrium was always maintained, the number of
WIMPs today would be infinitesimal.  However, the Universe is
not static, so equilibrium thermodynamics is not the entire story.

%
\begin{figure}[htbp]
\centerline{\psfig{file=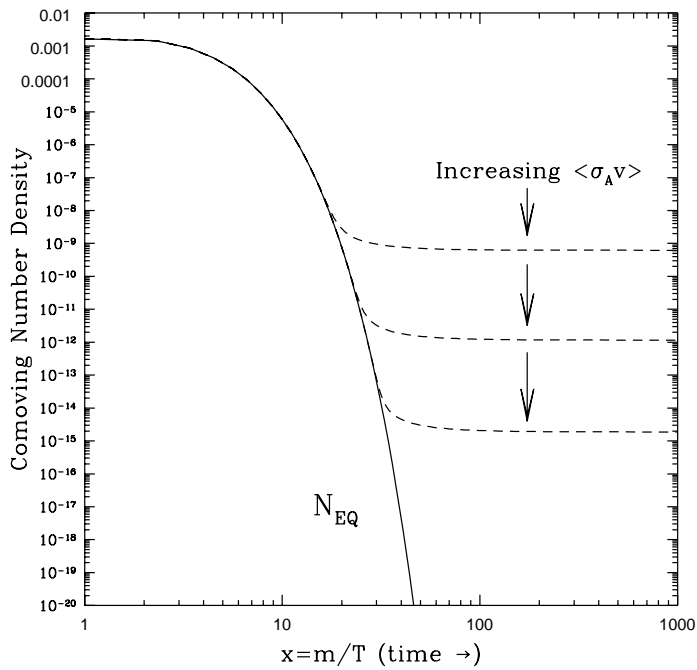,width=3.7in}}
\bigskip
\caption{\baselineskip=0.4cm
        Comoving number density of a WIMP in the early
	Universe.  The dashed curves are the actual abundance,
	and the solid curve is the equilibrium abundance.}
\label{YYY}
\end{figure}

At high temperatures ($T\gg m_\chi$), $\chi$'s are abundant and
rapidly converting to lighter particles and {\it vice versa}
($\chi\bar\chi\leftrightarrow l\bar l$, where $l\bar l$ are quark-antiquark and
lepton-antilepton pairs, and if $m_\chi$ is greater than the mass of the
gauge and/or Higgs bosons, $l\bar l$ could be gauge- and/or Higgs-boson
pairs as well).  Shortly after $T$ drops below $m_\chi$ the number
density of $\chi$'s drops exponentially, and the rate for annihilation of
$\chi$'s, $\Gamma=\VEV{\sigma v} n_\chi$---where $\VEV{\sigma v}$ is the
thermally averaged total cross section for annihilation of $\chi\bar\chi$
into lighter particles times relative velocity $v$---drops below 
the expansion rate, $\Gamma\la H$.  At this point, the $\chi$'s cease to
annihilate, they fall out of equilibrium, and a relic cosmological
abundance remains.

Fig.~\ref{YYY} shows numerical solutions to the Boltzmann
equation which determines the WIMP abundance.  The
equilibrium (solid line) and actual (dashed lines) abundances
per comoving volume are plotted as a
function of $x\equiv m_\chi/T$ (which increases with increasing time).
As the annihilation cross section is increased
the WIMPs stay in equilibrium longer, and we are left with a
smaller relic abundance.

An approximate solution to the Boltzmann equation yields the
following estimate for the current cosmological abundance of the
WIMP:
\begin{equation}
     \Omega_\chi h^2={m_\chi n_\chi \over \rho_c}\simeq
     \left({3\times 10^{-27}\,{\rm cm}^3 \, {\rm sec}^{-1} \over
     \sigma_A v} \right),
\label{eq:abundance}
\end{equation}
where $h$ is the Hubble constant in units of 100
km~sec$^{-1}$~Mpc$^{-1}$.  The result is to a first approximation
independent of the WIMP mass and is fixed primarily by its
annihilation cross section.

The WIMP velocities at freeze out are typically some appreciable
fraction of the speed of light.  Therefore, from
equation~(\ref{eq:abundance}), the WIMP will have a cosmological
abundance of order unity today if the annihilation cross section
is roughly $10^{-9}$ GeV$^{-2}$.  Curiously, this is the order
of magnitude one would expect from a typical electroweak cross
section, 
\begin{equation}
     \sigma_{\rm weak} \simeq {\alpha^2 \over m_{\rm weak}^2},
\end{equation}
where $\alpha \simeq {\cal O}(0.01)$ and $m_{\rm weak} \simeq
{\cal O}(100\, {\rm GeV})$.  The value of the cross section in
equation~(\ref{eq:abundance}) needed to provide $\Omega_\chi\sim1$
comes essentially from the age of the Universe.  However, there
is no {\it a priori} reason why this cross section should be of
the same order of magnitude as the cross section one would
expect for new particles with masses and interactions
characteristic of the electroweak scale.  In other words, why
should the age of the Universe have anything to do with
electroweak physics?  This ``coincidence'' suggests that if a
new, yet undiscovered, massive particle with electroweak
interactions exists, then it should have a relic density of
order unity and therefore provides a natural dark-matter
candidate.  This argument has been the driving force behind a
vast effort to detect WIMPs in the halo.

The first WIMPs considered were massive Dirac or Majorana
neutrinos with masses in the range of a few GeV to a few TeV.
(Due to the Yukawa coupling which gives a neutrino its mass, the
neutrino interactions become strong above a few TeV, and it no
longer remains a suitable WIMP candidate~\cite{unitarity}.)  LEP ruled out
neutrino masses below half the $Z^0$ mass.  Furthermore, heavier
Dirac neutrinos have been ruled out as the primary component of
the Galactic halo by direct-detection experiments (described
below)~\cite{heidelberg}, and heavier Majorana neutrinos have
been ruled out by indirect-detection
experiments~\cite{kamiokande} (also described below) over much 
of their mass range.  Therefore, Dirac neutrinos cannot comprise
the halo dark matter~\cite{griestsilk}; Majorana neutrinos can,
but only over a
small range of fairly large masses.  This was a major triumph
for experimental particle astrophysicists:\ the first
falsification of a dark-matter candidate.  However, theorists
were not too disappointed:  The stability of a fourth generation
neutrino had to be postulated {\it ad hoc}---it was not
guaranteed by some new symmetry.  So although heavy neutrinos
were plausible, they certainly were not very well-motivated from
the perspective of particle theory.

A much more promising WIMP candidate comes from supersymmetry
(SUSY)~\cite{jkg,haberkane}.  SUSY was
hypothesized in particle physics to cure the naturalness problem
with fundamental Higgs bosons at the electroweak scale.
Coupling-constant unification at the GUT scale seems to be
improved with SUSY, and it seems to be an essential ingredient
in theories which unify gravity with the other three fundamental
forces.

As another consequence, the existence of a new symmetry,
$R$-parity, in SUSY theories guarantees that the lightest
supersymmetric particle (LSP) is stable.
In the minimal supersymmetric extension of the
standard model (MSSM), the LSP is usually the neutralino, a linear
combination of the supersymmetric partners of the photon, $Z^0$,
and Higgs bosons.  (Another possibility is the sneutrino, but
these particles interact like neutrinos and have been ruled out
over most of the available mass range~\cite{sneutrino}.)  Given
a SUSY model, the cross section for
neutralino annihilation to lighter particles is straightforward,
so one can obtain the cosmological mass density.  The
mass scale of supersymmetry must be of order the weak scale to
cure the naturalness problem, and the neutralino will have only
electroweak interactions.  Therefore, it is to be expected that
the cosmological neutralino abundance is of order unity.  In
fact, with detailed calculations, one finds that the neutralino
abundance in a very broad class of supersymmetric extensions of
the standard model is near unity and can therefore account for
the dark matter in our halo~\cite{ellishag}.

If neutralinos reside in the halo, there are several avenues
toward detection~\cite{jkg}.  One of the most promising
techniques currently being pursued involves searches for the
${\cal O}(10\, {\rm keV})$ recoils produced by elastic scattering of
neutralinos from nuclei in low-background
detectors~\cite{witten,labdetectors}.  The idea here is simple.
A particle with mass $m_\chi\sim100$ GeV and electroweak-scale
interactions
will have a cross section for elastic scattering from a nucleus
which is $\sigma \sim 10^{-38}\,{\rm cm}^2$.  If the local halo
density is $\rho_0\simeq0.4$ GeV~cm$^{-3}$, and the particles
move with velocities $v\sim 300$ km~sec$^{-1}$, then the rate
for elastic scattering of these particles from, e.g., germanium
which has a mass $m_N \sim70$ GeV, will be $R \sim \rho_0
\sigma v / m_\chi/m_N \sim1$ event~kg$^{-1}$~yr$^{-1}$.  If a
$100$-GeV WIMP moving at $v/c\sim10^{-3}$ elastically scatters
with a nucleus of similar mass, it will impart a recoil energy
up to 100 keV to the nucleus.  Therefore, if we have 1 kg of
germanium, we expect to see roughly one nucleus per year
spontaneously recoil with an energy nearly 100 keV.

Of course, this is only a {\it very} rough calculation.  To do
the calculation more precisely, one needs to use a proper
neutralino-quark interaction, treat the QCD and
nuclear physics which takes you from a neutralino-quark
interaction to a neutralino-nucleus interaction, and integrate
over the WIMP velocity distribution.  Even if all of these
physical effects are included properly, there is still a
significant degree of uncertainty in the predicted event rates.
Although supersymmetry provides perhaps the most promising
dark-matter candidate (and solves numerous problems in particle
physics), it really provides little detailed predictive power.
In SUSY models, the standard-model particle spectrum is more
than doubled, and we really have no idea what the masses of all
these superpartners should be.  There are also couplings, mixing
angles, etc. Therefore, what theorists generally do is survey a
large set of models with masses and couplings within a
plausible range, and present results for relic abundances and
direct- and indirect-detection rates, usually as scatter plots
versus neutralino mass.  

After taking into account all the relevant physical effects and
surveying the plausible region of SUSY parameter space, one
generally finds that the predicted event rates seem to fall for
the most part between $10^{-4}$ to 10
events~kg$^{-1}$~day$^{-1}$, \cite{jkg} although again, there
may be models with higher or lower rates.  Current experimental
sensitivities in germanium detectors are around 10
events~kg$^{-1}$~day$^{-1}$. \cite{heidelberg}  To illustrate
future prospects, consider the CDMS experiment~\cite{cdms} which
expects to soon have a kg germanium detector with a background
rate of 1 event~day$^{-1}$.  After a one-year exposure, their
sensitivity would therefore be ${\cal O}(0.1\, {\rm
event~kg}^{-1}\,{\rm day}^{-1})$; this could be improved with
better background rejection.  Future detectors will achieve
better sensitivities, and it should be kept in mind that
numerous other target nuclei are being considered by other
groups.  However, it also seems clear that it will be quite a
while until a good fraction of the available SUSY parameter
space is probed.

Another strategy is observation of energetic neutrinos produced
by annihilation of neutralinos in the Sun and/or Earth in converted
proton-decay and astrophysical-neutrino detectors (such as
MACRO, Kamiokande, IMB, AMANDA, and NESTOR)~\cite{SOS}.  If, upon
passing through the Sun, a WIMP scatters elastically from a
nucleus therein to a velocity less than the escape velocity, it
will be gravitationally bound in the Sun.  This leads to a
significant enhancement in the density of WIMPs in the center of
the Sun---or by a similar mechanism, the Earth.  These WIMPs
will annihilate to, e.g., $c$, $b$, and/or $t$ quarks, and/or gauge and
Higgs bosons.  Among the decay products of these particles
will be energetic muon neutrinos which can escape from the
center of the Sun and/or Earth and be detected in neutrino
telescopes such as IMB, Kamiokande, MACRO, AMANDA, or NESTOR.
The energies of these muons will be typically 1/3 to 1/2 the
neutralino mass (e.g., 10s to 100s of GeV) so they will be much
more energetic---and therefore cannot be confused
with---ordinary solar neutrinos.  The signature of such a
neutrino would be the Cerenkov radiation emitted by an upward
muon produced by a charged-current interaction between the
neutrino and a nucleus in the rock below the detector.

The annihilation rate of these WIMPs is equal to the rate for
capture of these particles in the Sun.  This can be estimated in
order of magnitude by determining the rate at which halo WIMPs
elastically scatter from nuclei in the Sun.  The flux of
neutrinos at the Earth depends also on the Earth-Sun distance,
WIMP annihilation branching ratios, and the decay branching
ratios of the annihilation products.  The flux of upward muons
depends on the flux of neutrinos and the cross section for
production of muons, which depends on the square of the neutrino
energy.  As in the case of direct detection, the precise
prediction involves numerous factors from particle and nuclear
physics and astrophysics, and on the SUSY parameters.
When all these factors are taken into account, predictions for
the fluxes of such muons in SUSY models
seem to fall for the most part between $10^{-6}$ and 1
event~m$^{-2}$~yr$^{-1}$, \cite{jkg} although the numbers may be a bit
higher or lower in some models.  Presently, IMB and Kamiokande
constrain the flux of energetic neutrinos from the Sun to be
less than about 0.02 m$^{-2}$~yr$^{-1}$, \cite{kamiokande,imb}
and the Baksan limit is perhaps a factor-of-two
better~\cite{baksan}.  MACRO expects to be
able to improve on this sensitivity by perhaps an order of
magnitude.  Future detectors may be able to improve even
further.  For example, AMANDA expects to have an area of
roughly $10^4$ m$^2$, and a $10^6$-m$^2$
detector is being discussed.  However, it should be kept in
mind that without muon energy resolution, the sensitivity of
these detectors will not approach the inverse exposure; it will
be limited by the atmospheric-neutrino background.  If a
detector has good angular resolution, the signal-to-noise ratio
can be improved, and even moreso with energy resolution, so
sensitivities approaching the inverse exposure could be
achieved~\cite{joakim}.  Furthermore, ideas for neutrino detectors with
energy resolution are being discussed~\cite{wonyong}, although
at this point these appear likely to be in the somewhat-distant future.

With two promising avenues toward detection, it is natural to
inquire which is most promising.  Due to the abundance of
undetermined SUSY parameters and the complicated dependence of
event rates on these parameters, the answer to this question is
not entirely straightforward.
Generally, most theorists have just plugged in SUSY parameters
into the machinery which produces detection rates and plotted
results for direct and indirect detection.  However, another
approach is to compare, in a somewhat model-independent although
approximate fashion, the rates for direct and indirect
detection~\cite{jkg,taorich,bernard}.  The underlying
observation is that the rates for the 
two types of detection are both controlled primarily by the WIMP-nucleon
coupling.  One must then note that WIMPs generally undergo one
of two types of interaction with the nucleon: an axial-vector
interaction in which the WIMP couples to the nuclear spin
(which, for nuclei with nonzero angular momentum is roughly 1/2
and {\it not} the total angular momentum), and a scalar
interaction in which the WIMP couples to the total mass of the
nucleus.  The direct-detection rate depends on the WIMP-nucleon
interaction strength and on the WIMP mass.  On the other hand,
indirect-detection rates will have an additional dependence on
the energy spectrum of neutrinos from WIMP annihilation.  By
surveying the various possible neutrino energy spectra, one
finds that for a given neutralino mass and annihilation rate in
the Sun, the largest upward-muon flux is roughly three times as
large as the smallest~\cite{bernard}.  So even if we assume the
neutralino-nucleus interaction is purely scalar or purely
axial-vector, there will still be a residual model-dependence of
a factor of three when comparing direct- and indirect-detection
rates.

For example, for scalar-coupled WIMPs, the event rate in a kg
germanium detector will be
equivalent to the event rate in a $(2-6)\times 10^6$ m$^2$
neutrino detector for 10-GeV WIMPs and $(3-5)\times10^4$ m$^2$
for TeV WIMPs~\cite{bernard}.  Therefore, the relative
sensitivity of indirect detection when compared with the
direct-detection sensitivity increases with mass.
The bottom line of such an analysis seems to be that
direct-detection experiments will be more sensitive to
neutralinos with scalar interactions with nuclei, although
very-large neutrino telescopes may achieve comparable
sensitivities at larger WIMP masses.  This should
come as no surprise given the fact that direct-detection
experiments rule out Dirac neutrinos~\cite{heidelberg}, which
have scalar-like interactions, far more effectively than
do indirect-detection experiments~\cite{bernard}.

Generically, the sensitivity of
indirect searches (relative to direct searches) should be better
for WIMPs with axial-vector interactions, since the Sun is
composed primarily of nuclei with spin (i.e., protons).
However, a comparison of direct-
and indirect-detection rates is a bit more difficult for
axially-coupled WIMPs, since the nuclear-physics uncertainties
in the neutralino-nuclear cross section are much greater, and
the spin distribution of each target nucleus must be modeled.
Still, in a careful analysis, Rich and Tao found that in 1994,
the existing sensitivity of energetic-neutrino searches to
axially-coupled WIMPs greatly exceeded the sensitivities of
direct-detection experiments~\cite{taorich}.

To see how the situation may change with future
detectors, let us consider a specific axially-coupled
dark-matter candidate, the light Higgsino recently put forward by
Kane and Wells~\cite{kanewells}.  In order to explain the
anomalous CDF $ee\gamma
\gamma + \slashchar{E}_T$~\cite{CDF}, the $Z\rightarrow b\bar b$ anomaly,
and the dark matter, this Higgsino must have a mass between
30--40 GeV.  Furthermore, the coupling of this Higgsino to
quarks and leptons is due primarily to $Z^0$ exchange with a
coupling proportional to $\cos 2\beta$, where $\tan\beta$ is the
usual ratio of Higgs vacuum expectation values in
supersymmetric models.  Therefore, the usually messy cross
sections one deals with in a general MSSM simplify for this
candidate, and the cross sections needed for the cosmology of
this Higgsino depend only on the two parameters $m_\chi$ and
$\cos2\beta$.  Furthermore, since the neutralino-quark
interaction is due only to $Z^0$ exchange, this Higgsino will
have only axial-vector interactions with nuclei.

The Earth is composed primarily of spinless nuclei, so WIMPs
with axial-vector interactions will not be captured in the Earth,
and we expect no neutrinos from WIMP annihilation therein.
However, most of the mass in the Sun is composed of nuclei with
spin (i.e., protons).  The flux of upward muons induced by
neutrinos from
annihilation of these light Higgsinos would be $\Gamma_{\rm
det}\simeq 2.7\times10^{-2}\, {\rm m}^{-2}\, {\rm yr}^{-1}\,
\cos^2 2\beta$. \cite{katie}  On the other hand, the rate for scattering from
$^{73}$Ge is $R\simeq 300\, \cos^2 2\beta\, {\rm kg}^{-1}\,
{\rm yr}^{-1}$. \cite{kanewells,katie}  For illustration, in
addition to their kg of
natural germanium, the CDMS experiment also plans to
run with 0.5 kg of (almost) purified $^{73}$Ge.  With a
background event rate of roughly one event~kg$^{-1}$~day$^{-1}$,
after one year, the $3\sigma$ sensitivity of the experiment will
be roughly 80 kg$^{-1}$~yr$^{-1}$.  Comparing the predictions
for direct and indirect detection of this axially-coupled WIMP,
we see that the enriched-$^{73}$Ge sensitivity should improve on
the {\it current} 
limit to the upward-muon flux ($0.02$ m$^{-2}$ yr$^{-1}$)
roughly by a factor of 4.  When we compare this with the forecasted
factor-of-ten improvement expected in MACRO, it appears that the
sensitivity of indirect-detection experiments looks more
promising.  Before
drawing any conclusions, however, it should be noted that the
sensitivity in detectors with other nuclei with spin may be
significantly better.  On the other hand, the sensitivity of
neutrino searches increases relative to direct-detection
experiments for larger WIMP masses.  It therefore seems at this
point that the two schemes will be competitive for detection of
light axially-coupled WIMPs, but the neutrino telescopes may
have an advantage in probing larger masses.

A common question is whether
theoretical considerations favor a WIMP which has predominantly
scalar interactions or whether they favor axial-vector
couplings.  Unfortunately, there is no
simple answer.  When detection of supersymmetric dark matter was
initially considered, it seemed that the neutralino in most
models would have predominantly axial-vector interactions.  It
was then noted that in some fraction of models where the
neutralino was a mixture of Higgsino and gaugino, there could be
some significant scalar coupling as well~\cite{kim}.  As it became evident
that the top quark had to be quite heavy, it was realized that
nondegenerate squark masses would give rise to scalar couplings
in most models~\cite{drees}.  However, there are still large regions of
supersymmetric parameter space where the neutralino has
primarily axial-vector interactions, and in fact, the Kane-Wells
Higgsino candidate has primarily axial-vector interactions.  The
bottom line is that theory cannot currently reliable say which
type of interaction the WIMP is likely to have, so
experiments should continue to try to target both.

\section{Discussion}

There has been no shortage of proposed solutions since the
advent of the dark-matter problem nearly seventy years ago.
Moreover, there has been an explosion in the number of exotic
solutions proposed in recent decades.  However, among the
numerous proposals, only WIMPs and axions have really survived
extended theoretical scrutiny for close to twenty years.  Unlike
other proposed dark-matter candidates, neither WIMPs nor axions
really require any exotic mechanisms in the early Universe to
guarantee a relic density of order unity---the only speculation
is that they exist, and if they do, they
have a relic density near unity.  Furthermore, it is important
to keep in mind that these particles were invented by particle
theorists to solve problems with the standard model and only
later was it realized that they were dark-matter candidates;
they were not introduced to solve the dark-matter problem.

Axions and WIMPs have not only intrigued theorists; a large
community of experimentalists have devoted themselves to finding
these particles.  However, it should also be emphasized that
although very attractive, these are still speculative ideas.
There is still no direct evidence in accelerator experiments or
otherwise for the existence of axions or of supersymmetry.  The
dark matter could be composed of something completely
different.  However, as argued here, the evidence for the
existence of nonbaryonic dark matter is indeed extremely
compelling, and the two particles discussed here provide our
most promising candidates.  Although it provides an enormous
experimental challenge, it is clear that discovery of particle
dark matter would be truly revolutionary for both particle
physics and cosmology.

\section*{Acknowledgments}
This work was supported by the D.O.E. under
contract DEFG02-92-ER 40699, NASA under NAG5-3091, and the
Alfred P. Sloan Foundation.

\end{document}